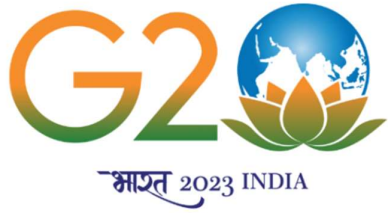
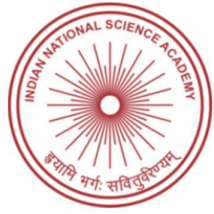
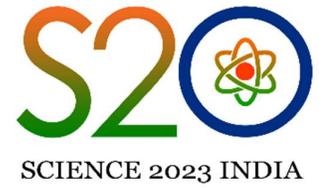

# Policy Brief

## LONG TERM SPACE DATA AND INFORMATICS NEEDS

Discussed and drafted during S20 Policy Webinar on Astroinformatics for Sustainable Development held on 6-7 July 2023

Contributors: S. Bradley Cenko, Daniel Crichton, Richard J. Doyle, Seetha Somasundaram, Giuseppe Longo, Laurent Eyer, Pranav Sharma, Ashish Mahabal

**Introduction**
Persistent space data gathering, retention, transmission, and analysis play a pivotal role in deepening our grasp of the Universe and fostering the achievement of global sustainable development goals. Long-term data storage and curation is crucial not only to make the wide range of burgeoning data sets available to the global science community, but also to stabilize those data sets, enabling new science in the future to analyse long-term trends over unprecedented time spans.

In addition to this, over the long-term, the imperative to store all data on the ground should be ameliorated by use of space-based data stores—maintained and seen to be as reliable as any other data archive. This concept is sometimes referred to as Memory of the Sky. Storing the data must be accompanied by the ability to analyse them. Several concepts covered below acknowledge roots and inspiration based in the Virtual Observatory effort.

Within this policy document, we delve into the complexities surrounding the long-term utilization of space data and informatics, shedding light on the challenges and opportunities inherent in this endeavour. Further, we present a series of pragmatic recommendations designed to address these challenges proactively.

**Challenges**
Long-term space data collection and storage present several challenges, including data transmission, management, and analysis. These challenges can be addressed by developing and implementing new technologies and data processing methods.

**Recommendations**
1. Develop advanced **data processing and analysis** techniques: Develop and implement new data processing techniques such as machine learning algorithms and predictive models to help manage and analyse the vast amounts of data collected from space. As part of an Open Science approach, the algorithms and software supporting machine learning, other analytics, and models, should be made available and be well-documented to support reproducibility of results.

    **Analytics** should be applied within the space-based system, as well as on the ground, with objectives such as: removing poor quality or compromised data, prioritizing data based on intermediate assessment to manage overall downlink resources, and, as appropriate, enable discovery and characterization of new phenomena. Space-based analytics are subject to computing, networking, and data storage constraints.

    **Data provenance** is a critical concern. To ensure comprehensive traceability, it is important to keep track of input and output, and configuration parameters, as well as the reference to the software used, along with its version. This approach allows for a complete record of all relevant information related to the analysis process. This process can be divided into two main components: the technical pipeline and the human-oriented aspect.

2. Encourage **international cooperation**: Promote international cooperation and collaboration to help share resources and expertise, facilitate data sharing, and ensure that data is available to researchers worldwide. This enables the actual use of data—which is the primary aim of taking the data in the first place. Most science mission data is available for scientists to download. However, there are opportunities for better ways of sharing—possibly reducing data at some sites.

Large-scale missions necessitate extensive international collaborations. This requirement arises from the immense scope of such projects, where no single institute possesses the entirety of experience and expertise necessary to fully address the processing and analysis requirements.

The global proliferation of Cube/SmallSats for capabilities in low-Earth orbit is making cooperation, coordination and ultimately, space traffic management increasingly imperative.

3. Adopt common standards/ Develop standardized data formats and metadata: Encourage the development, use and when possible, adoption of standardized data formats and metadata, to help facilitate data sharing and interoperability among different data archives and scientific communities. The International Virtual Observatory Alliance (IVOA) has done important work to date in this area within the astronomy community. Standards developed by NASA's Planetary Data System (PDS) for planetary data are being used internationally (e.g. by Indian Science Missions).

Concerning sky surveys, interoperability is important: the ability to merge data, define the epoch of a survey, and correctly propagate proper motion, as well as understanding the general properties of each survey. For example, for photometric surveys, often different bands and photometric systems are utilized. For useful comparisons, or merging data, it is important to transform bands to a common base, which depends on the scientific goal. While some steps have been taken in this direction, the steps are generally not uniform or easily transferable to other workflows.

4. Invest in the development and maintenance of robust **data storage infrastructure**—often cloud solutions—to ensure the long-term preservation and accessibility of data[1]. The usage of computing accelerators such as GPUs (Graphics Processing Units) or FPGAs (Field Programmable Gate Arrays), both good options to shorten processing times. The cost-benefit balance needs to be evaluated. Computation close to the data is in general good. In case of space-based data it can be even better to reduce data volume to be transferred.

---

[1] ESA has a Science Archives Long-Term Strategy (last issued in 2018) to guarantee the availability of science data to future generations beyond the 10-20+ years horizon. The strategy is based on substantial consultation with the international scientific community and internal technical and scientific staff, and is based on three main pillars:
   a) Maximize the potential of scientific return of the datasets, achieved by generation and preservation of high-quality multi-mission, multi-wavelength, multi-epoch, and multi-instrument data services in ways easily usable by scientists (enabling the FAIR principle, for Findable, Accessible, Interoperable and Reusable). An example of such services is the ESASky platform, which provides access to most professional astronomical data produced world-wide from a simple-to-use and intuitive web application.
   b) Ensure long-term preservation of data, knowledge, and associated access services. This objective requires technical evolution of the software packages and interfaces used to process data from raw data to high-level science products. A robust technology-aware program is needed to ensure long-term maintenance. ESA is currently working on a science platform, called ESA Datalabs, that will enable the hosting of public data alongside the processing algorithms, in a cloud-based infrastructure with easy access to the world-wide community.
   c) Treat the science data handling, from the instrument to the scientific publication, as an integral part of the project plan to develop the various missions, surveys or telescopes. This ensures continued funding and integration of the archives with science operations, and guarantees a smoother transition into the legacy phase, once the telescope, survey or mission completes and specific funding for archiving ceases to exist.

5. Promote **education and outreach** efforts to raise awareness about long-term space data and informatics needs and their importance for scientific research, global development, and STEM education. Opportunity exists to harness the educational potential of utilizing scientific data for public outreach and instruction, which remains largely untapped. The number of examples of outreach and citizen science projects is increasing (recent GaiaVari, Planet Hunters TESS, etc.), applications on smartphones (ZTF's ZARTH), but current opportunities still only scratch the surface.

6. Foster **partnerships** among academia, industry, and government agencies to leverage expertise and resources, and develop new technologies and infrastructure to support long-term space data collection and analysis.

7. Data science must address needs across the full **space data lifecycle**, from point-of-origin at the space platform, to available curated data at distributed ground-based archives. This span includes structuring and labeling data for AI/ML; future needs to be able to capture and share training sets; ability to capture/share software models for AI/ML.

    Relevant full-lifecycle activities include: performing original processing at the sensor or instrument; making choices at the collection point about which data to keep; improving resource efficiencies to enable moving the most data; anticipating the need to work across multiple data sources; increasing computing capability at the data to generate products; increasing the scale and integration of distributed archives; applying visualization techniques to increase data understanding; applying machine learning and statistics to enable data understanding; and creating analytics services across massive, distributed data.

    Success metrics and associated measures fall into four broad areas:
    1) Data Providers (early lifecycle), e.g., variety, velocity, and volume
    2) Data System (middle lifecycle), e.g., effectiveness, flexibility, reliability and sustainability
    3) Data Users (late lifecycle), e.g., accessibility, reproducibility, utility, yield
    4) Data Architecture (end-to-end), e.g., integrity, tradeability, scalability.

**Conclusion**
Long-term space data and informatics needs present significant challenges, but also offer enormous potential for advancing scientific research, supporting global development goals, and promoting STEM education through sharing of e.g., Jupyter notebooks and other analytical tools that can integrate with massive archives. Through the recommendations presented here, we can help address these challenges and ensure that long-term space data remains a valuable resource for future generations of researchers and scientists.

**S20 Co-Chair**: Ashutosh Sharma, Indian National Science Academy
**INSA S20 Coordination Chair:** Narinder Mehra, Indian National Science Academy

**Contributors**
S. Bradley Cenko, NASA's Goddard Space Flight Center, USA
Daniel Crichton, Leader, Center for Data Science and Technology, Jet Propulsion Laboratory, USA
Richard J. Doyle, Program Manager for Information and Data Science (retired), currently, Technical Consultant, NASA Jet Propulsion Laboratory, USA
Seetha Somasundaram, Former Director of Space Science Programme Office, ISRO, currently Emeritus Scientist (Honorary), Raman Research Institute, Bangalore, India
Guissepe Longo, Università degli Studi di Napoli Federico II, Italy
Laurent Eyer, University of Geneva, Switzerland
Pranav Sharma, Indian National Science Academy, India
Ashish Mahabal, California Institute of Technology, USA